\documentstyle[twocolumn,amssymb,eqsecnum,aps]{revtex}

\begin{document}
\draft
\title{Peltier effect induced longitudinal resistivity of ideal 2D electron(hole)
gas in strong magnetic field}
\author{M. V. Cheremisin$^{*}$}
\address{A.F.Ioffe Physical-Technical Institute,St.Petersburg, Russia}
\date{\today}
\maketitle

\begin{abstract}
We demonstrate that in strong quantum limit the thermoelectric Peltier
effect could define the longitudinal resistivity of dissipationless
two-dimensional electron(hole) gas. The current results in heating(cooling)
at first(second) Hall bar sample contact due to Peltier effect. At small
current the contacts temperatures are different, the temperature gradient is
linear on current. The voltage swing downstream the current is proportional
to Peltier effect induced thermopower. As a result, nonzero longitudinal
resistivity is measured in experiment. The above effect could exist in 3D
case.
\end{abstract}


The understanding of electronic transport in solid states subjected into
high magnetic fields has undergone a profound revision after the discovery
of integral$^{\text{\cite{Klitzing}}}$ and fractional quantum Hall effect$^{%
\text{\cite{Tsui}}}$ for two-dimensional electron gas( 2DEG ). Both
phenomena manifest extraordinary transport behavior as the temperature
approaches to zero. The Hall resistivity, $\rho _{xy}$ ,is quantized to $%
h/ie^{2}$ with $i$ being either an integer or rational fraction while
longitudinal magnetoresistivity, $\rho _{xx}$ , vanishes.

The main goal of this paper is to present a thermodynamic approach regarding
to IQHE problem. The crucial point is that for dissipationless 2DEG the
on-diagonal components of resistivity tensor could arise from combined
thermoelectric Peltier and Seebeck effects.$^{\text{\cite{Kirby},\cite
{Cheremisin}}}$ The current results in heating(cooling) at first(second)
Hall bar sample contact due to Peltier effect. The contacts temperatures are
different, the downstream temperature gradient is linear on current. The
voltage swing downstream the current is equal to Peltier effect induced
thermopower which is linear on current. As a result, nonzero longitudinal
resistivity is measured in experiment. This value could be erroneously
identified as 2D resistivity which is equal, in contrast, to zero for
dissipationless 2DEG in high magnetic field.

Let us consider 2D electron gas in x-y plane(Fig.\ref{Fig.1}) subjected into
perpendicular magnetic field $B=B_{z}$. The 2DEG structure ( MOS, quantum
well etc.) is arbitrary, the electrons assumed to occupy the first size
quantization subband. Neglecting spin splitting, the energy spectrum is $%
\varepsilon _{n}=\hbar \omega _{c}(n+1/2),$where $n=0,1$... is Landau level
(LL) number, $\omega _{c}=eB/mc$ is the cyclotron frequency, $m$ is the
electron effective mass. For simplicity, we omit LL broadening, thus, the
LLs density of states is $\Gamma =eB/hc$. We will further restrict ourselves
within strong quantum limit $\hbar \omega _{c}>>kT$. If the case, one could
neglect any random short(long) range impurity scattering$^{\text{\cite
{Baskin}}}$, hence, consider dissipationless( ideal ) 2DEG. Then, for actual 
$T\rightarrow 0$ limit we disregard any phonon related effects($\sim T^{3}$)
in both the bulk and 2DEG.

According to conventional thermodynamics, at thermodynamic equilibrium the
chemical potential, $\mu $, for the system conductors+2DEG is constant.
Moreover, at small perturbation of thermodynamic equilibrium due to current
the chemical potential remains constant. Thus, we conclude that 2D electron
gas is, in fact, non-isolated. We argue that an external reservoir of
electrons, if exists, could provide the pining of 2DEG chemical potential.
Once supposed that $\mu $ is fixed, it could be shown that 2DEG
concentration, $N$, changes with magnetic field. Indeed, at $T\rightarrow 0$
one needs $N=i\Gamma $ electrons in order to occupy $i$ levels, therefore,
the transverse conductivity is given $\sigma _{yx}=Nec/H=ie^{2}/h.$
Accordingly, the external reservoir must offer the possibility of changing $%
N $ over sufficiently wide limits in order to achieve observable widths of $%
\sigma _{yx}$ plateau. The above idea was put forward first in Ref.\cite
{Barraf} where ionized donors at a right distance to heterojunction assumed
to serve as such reservoir. Then, the surface states shown in Ref.\cite
{Konstantinov} could play the same role in MOS structures. At a moment,
above scenario is abandoned because of Ref. \cite{Ebert} in which was
underlined insufficient strength of reservoir in both models. Nevertheless,
the experimental results found in Ref.\cite{Nizhankovskii} are consistent
with reservoir conception. We argue that thermodynamics based argumentation
done above is general, thus, we use reservoir idea assuming that 2DEG
chemical potential is fixed. The detailed analysis of actual reservoir will
be done elsewhere.

Let us consider Hall bar geometry sample(Fig.\ref{Fig.1}) under dc current
carrying conditions. The sample is connected by means of two identical extra
leads to the current source (not shown). Both contacts are ohmic ones. The
voltage is measured between the open ends (''c'' and ''d'') kept at fixed
temperature. The 2D structure is placed into the sample chamber ( not shown
) with the mean temperature $T_{0}$. According to our basic assumption the
temperature gradient, $\nabla _{x}T$ , is nonzero downstream the current. In
general, for bounded topology sample the macroscopic current, ${\bf j}$ ,%
\hspace{0in} and the energy flux, ${\bf q}$ , densities are given$^{\cite
{Streda}}$

\begin{eqnarray}
{\bf j} &=&{\bf \sigma }^{\symbol{94}}({\bf E-}\alpha \nabla T),  \eqnum{1}
\label{current+heat} \\
{\bf q} &=&\left( \alpha T-\frac{\zeta }{e}\right) {\bf j}-{\bf \varkappa }^{%
\symbol{94}}\nabla T.  \nonumber
\end{eqnarray}
Here, ${\bf E}=\frac{1}{e}\nabla \zeta $ is the electric field, $\zeta =\mu
-e\varphi $ is the electrochemical potential, $\alpha $ is the 2DEG
thermopower.$^{\text{\cite{Girvin},\cite{Zelenin}}}$ For dissipationless
2DEG in strong quantum limit $\hbar \omega _{c}>>kT$ the conductivity
tensor, ${\bf \sigma }^{\symbol{94}}$, is purely off-diagonal. Hence, $%
\sigma _{yx}=1/\rho _{yx}=e^{2}/h\sum\limits_{n}f(\varepsilon _{n})$ is the
transverse conductance at finite temperature, $f(\varepsilon )$ is the Fermi
function. Then, in Eq.(\ref{current+heat}) ${\bf \varkappa }^{\symbol{94}%
}=(L-\alpha ^{2})T{\bf \sigma }^{\symbol{94}}$ is the electron related
thermal conductivity, $L=\frac{\pi ^{2}k^{2}}{3e^{2}}$ is the Lorentz
number. Note, in Eq.(\ref{current+heat}) the diamagnetic surface currents$^{%
\text{\cite{Obraztzov}}}$ are yet accounted, thus, both equations satisfy
Einstein and Onsager relationships. We emphasize that Eq.(\ref{current+heat}%
) could be derived(see Appendix A) by means of drift approach developed in
Ref.\cite{Zyryanov1},\cite{Zyryanov2} for 3D case.

With the help of Eq.(\ref{current+heat}), the current components are: 
\begin{eqnarray}
j_{x} &=&\sigma _{xy}E_{y},  \eqnum{2}  \label{j-components} \\
j_{y} &=&\sigma _{yx}(E_{x}-\alpha \nabla _{x}T)=0.  \nonumber
\end{eqnarray}
The longitudinal current, $j=j_{x}$, represents the flux of electrons in
crossed fields with the drift velocity $v_{dr}=cE_{y}/B$ , where $E_{y}$ is
the Hall electric field. The transverse electron flow caused by longitudinal
electric field, $E_{x}$, is mutually compensated by the one because of
temperature gradient, hence $j_{y}=0$.

We now find out the longitudinal temperature gradient $\nabla _{x}T$ caused
by Peltier effect. We recall that Peltier heat is generated by the current
crossing the contact of two different conductors. At the contact ( let say
''a'' in Fig.(\ref{Fig.1})) the temperature, the electrochemical potential $%
\zeta $, the normal components of the total current, $I$, and the total
energy flux are all continuous. Note, there exists the difference, $\Delta
\alpha =\alpha _{1}-\alpha _{2}$, of the conductor and 2D gas thermopower
respectively. For $\Delta \alpha >0$ the charge intersecting the contact
''a'' gains the energy $e\Delta \alpha T_{a}$. Consequently, $Q_{a}=I\Delta
\alpha T_{a}$ is the amount of Peltier heat evolved per unit time in the
contact ''a''. For $\Delta \alpha >0$ and current direction shown in Fig.(%
\ref{Fig.1}) the contact ''a'' is heated while the contact ''b'' is cooled.
The contacts are at different temperatures: $\Delta T=T_{a}-T_{b}>0$. At $%
I\rightarrow 0$ the temperature gradient is small, hence, $%
T_{a},T_{b}\approx T_{0}$. Under above conditions the thermopowers are
nearly constants, thus, one could neglect Thomson heat, $Q_{T}\sim IT\nabla
\alpha $, downstream the current. Actually, the amount Peltier heat evolved
at the contact ''a'' is equal to one absorbed at the contact ''b''. Note, at 
$T_{0}\rightarrow 0$ the external cooling of 2DEG due to phonon thermal
conductivity($\sim T^{3}$) is suppressed. Hence, we assume adiabatic cooling
conditions of 2DEG.

We remind that the total energy flux is continuous at both ''a'' and ''b''
contacts. Using Eq.(\ref{current+heat}) one obtains 
\begin{eqnarray}
-\varkappa _{yx}\left[ \frac{dT}{dx}\right] _{a} &=&-jT_{a}(\alpha
_{2}-\alpha _{1}),  \eqnum{3}  \label{conditions} \\
\varkappa _{yx}\left[ \frac{dT}{dx}\right] _{b} &=&-jT_{b}(\alpha
_{1}-\alpha _{2}).  \nonumber
\end{eqnarray}
Here, we take into account that current is known to enter and leave the
sample at two diagonally opposite corners ( Fig.\ref{Fig.1}, insert a).
Consequently, ${\bf j\parallel q}_{b}=-\varkappa _{yx}\nabla _{x}T$ at the
contacts. In Eq.(\ref{conditions}) we neglect thermal transfer within the
leads keeping in mind 2DEG adiabatic cooling conditions. Using Eq.(\ref
{conditions}), the temperature gradient is given

\begin{equation}
\frac{dT}{dx}=\frac{-j\Delta \alpha T}{\varkappa _{yx}},  \eqnum{4}
\label{gradient}
\end{equation}
thus, linear on current. The voltage swing , $U$ , measured between the ends
''c'' and ''d'' yields 
\begin{equation}
U=\int\limits_{c}^{d}E_{x}dx=\int\limits_{c}^{d}\alpha dT=\Delta \alpha
(T_{a}-T_{b}).  \eqnum{5}  \label{temf}
\end{equation}
Here, we disregard the conductors resistances. Finally, the Peltier effect
related resistivity, $\rho =U/jl$ , could be written as follows 
\begin{equation}
\rho =\frac{\alpha ^{2}}{\sigma _{yx}(L-\alpha ^{2})},  \eqnum{6}
\label{CorrectionB}
\end{equation}
where we tale into account that for actual case of metallic leads $\Delta
\alpha \simeq -\alpha _{2}=-\alpha $, then $l$ is the sample length. Note,
Eq.(\ref{CorrectionB}) is valid for standard 4-probe Hall bar measurements,
2DHG ( Fig.\ref{Fig.1}, insert b) and in 3D case.

We recall, in strong quantum limit 2DEG thermopower is the universal
thermodynamic quantity which is proportional to entropy per one electron:$^{%
\cite{Girvin},\cite{Zelenin}}$ 
\begin{equation}
\alpha =-\frac{s}{e}=\frac{1}{eT}\frac{\sum\limits_{n}\int\limits_{%
\varepsilon _{n}}^{\infty }d\varepsilon (\varepsilon -\mu )f^{\prime
}(\varepsilon )}{\sum\limits_{n}f(\varepsilon _{n})}.  \eqnum{7}
\label{thermopower}
\end{equation}
Here, $\Omega (\mu ,T,B)=-kT\cdot \Gamma \sum\limits_{n}\ln (1+\exp ((\mu
-\varepsilon _{n})/kT))$ is the thermodynamic potential density. For given $%
\mu $ thermopower is the universal function of reduced temperature, $\xi
=kT/\mu $, and dimensionless magnetic field $\hbar \omega _{c}/\mu =\nu
^{-1} $, where $\nu $ is so-called filling factor. Using Eq.(\ref
{thermopower}) we derived (see Appendix B) the asymptotic formulae for 2DEG
thermopower which is valid within low temperature and low magnetic field
limit $\nu ^{-1},\xi <1$.

Let us now discuss the features of resistivity given by Eq.(\ref{CorrectionB}%
). We underline that $\rho $ could be expressed in fundamental units $%
h/e^{2} $ because $\Delta \alpha \simeq -\alpha \sim k/e$, hence, $\rho \sim
1/\sigma _{yx}$. Then, according to Eq.(\ref{CorrectionB},\ref{thermopower}%
), $\rho $ is the universal function of $\nu ,\xi $. In Fig.(\ref{Fig.2})
the resistivity $\rho $ and off-diagonal component $\rho _{yx}$ vs $\nu
^{-1}\sim B$ are displayed. When the chemical potential lies between the
LLs( $\nu =1,2..$ ), resistivity $\rho $ is thermally activated with an
activation energy being of the order of magnetic energy. Therefore, near LLs 
$\rho (\nu ^{-1})$ dependence has a series of large peaks. At $T\rightarrow
0 $ the magnitudes of thermopower and $\rho $ at half-filled LL ( $\nu
_{c}=1/2,3/2...$ ) approach the universal values $\alpha _{c}=\frac{k}{e}%
\frac{\ln 2}{\nu _{c}}$ and $\rho ^{c}=\frac{h}{e^{2}}\frac{\gamma }{%
1-\gamma }\nu _{c}^{-1}(\gamma =\alpha _{c}^{2}/L)$ independent of the
temperature, electron effective mass etc. Note, $\gamma <1$ for fillings $%
\nu _{c}>\sqrt{3}(\ln 2/\pi )=0.38$ , i.e. up to half-filled first LL. Then,
for higher magnetic fields $\gamma \simeq 1$ and our basic approach $%
T_{a},T_{b}\approx T_{0}$ becomes invalid. We underline that at critical
fillings $\nu _{c}$ the transverse resistivity approaches the universal
value, $\rho _{yx}^{c}=$ $\frac{h}{e^{2}}\nu _{c}^{-1}$, as well$.$ Using
Eq.(\ref{CorrectionB}) one could demonstrate that for $\gamma <<1$ in the
vicinity of critical fillings $\triangle \nu =\nu -\nu _{c}$ the thermal
correction could be presented in the form $\rho =\rho ^{c}\exp (-\triangle
\nu /\nu _{0}),$where $\triangle \nu /\nu _{0}<<1$. Here, $\nu _{0}(\xi
)=4\nu _{c}^{2}\xi /3$ is the sample and temperature dependent logarithmic
slope. For typical GaAs based 2D electron gas $(n=10^{11}$cm$^{-2}$ at $B=$0$%
)$ at $\nu _{c}=3/2$ one obtains $\nu _{0}(T)=0.11\cdot T($K$)$. We stress
that above universal $\rho ^{c}$, $\rho _{yx}^{c}$ values could be
attributed to so-called ''QH transition points'' discussed in press$^{\cite
{Shahar},\cite{Coleridge}}$. To test this, in Fig.\ref{Fig.3} the detailed
dependences $\rho (\nu ),\rho _{yx}(\nu )$ in the vicinity of $\nu _{c}=3/2$
for different temperatures $\xi $ are presented. Evidently, this set of
curves could be collapsed into a single curve since $\rho ,\rho _{yx}$ are
universal functions of $\xi $. We argue, the collapse is governed by a
single parameter which could be but linear on temperature. Experimentally,
in Ref.\cite{Shahar},\cite{Coleridge} was found that both $\rho (\nu ),\rho
_{yx}(\nu )$ dependences are well collapsed. Then, the logarithmic slope $%
\nu _{0}(T)$ found experimentally is linear on $T$ well down to the lowest
temperatures accessed. It should be noticed, however, that the heights of $%
\rho $ peaks are lower that ones found in experiment. The above discrepancy
could be attributed, for example, to phonon drag related enhancement of
thermopower which is omitted in our simple approach.

Let us find out in standard fashion ${\bf \sigma }^{\symbol{94}}=({\bf \rho }%
^{\symbol{94}})^{-1}$ \ the quantities, $\sigma _{xx}^{*}{},\sigma
_{yx}^{*}{}$ , called the transverse and the longitudinal conductivities
respectively: 
\begin{equation}
\sigma _{xx}^{*}=\sigma _{yx}\left( \frac{s}{1+s^{2}}\right) ,\sigma
_{yx}^{*}=\sigma _{yx}\left( \frac{1}{1+s^{2}}\right) ,  \eqnum{8}
\label{conditivity}
\end{equation}
where $s=\gamma /(1-\gamma )$ is the dimensionless parameter. Using Eq.(\ref
{conditivity}) one could derive the relationship $(\sigma
_{xx}^{*})^{2}+(\sigma _{yx}^{*})^{2}=\sigma _{yx}^{*}\sigma _{yx}$ instead
well known semicircle relation. In Fig.\ref{Fig.3} the $\sigma
_{xx}^{*}{},\sigma _{yx}^{*}$ dependences are plotted in the vicinity $\nu
_{c}=3/2$. Evidently, these curves could be collapsed into a single plot
well as $\rho ,\rho _{yx}$( see above discussion). This result is consistent
with experiments.$^{\text{\cite{Shahar},\cite{Coleridge}}}$ We emphasize, in
practice $\sigma _{xx}^{*}{},\sigma _{yx}^{*}$ are always derived from Hall
bar sample resistivity data. In fact, experimentally the interconnection
between the components of both resistivity and conductivity tensors is not
yet established.$^{\cite{Dolgopolov}}$ The only within QH plateau the direct
Corbino topology measurements$^{\cite{Dolgopolov}}$ of transverse
conductivity demonstrated that $\sigma _{yx}^{*}=\sigma _{yx}=1/\rho _{yx}$.

We now find out the temperature gradient established in Hall bar sample at,
for example, half-filled LL $\nu _{c}=5/2$. According to Eq.(\ref
{thermopower}) one obtains $\alpha =k\ln 2/(e\nu _{c})\approx 0.28k/e$, then 
$\rho _{yx}=2/5h/e^{2}$, thus $\gamma =0.02.$ Assuming sample width $d=1$mm,
length $l=3$mm and typical dc current $I=10$nA one could find out the
temperature gradient given by Eq.(\ref{gradient}) as $\nabla _{x}T\simeq
\alpha I\rho _{yx}/(Ld)\approx 22$ mK/mm. At helium temperatures $%
(T_{a}-T_{b})/T_{0}\approx 0.02<<1$, hence, our approach is valid. We
underline that the temperature gradient is proportional to the ratio $I/d$,
thus, could be the reason of current and sample width dependent scaling
observed in \cite{CurScaling},\cite{WidthScaling} respectively. For example,
let the Hall probes(Fig.(\ref{Fig.1}), insert a) are shifted by length, $%
x_{0}$ , with respect to middle of the sample. The local temperature
established between the probes is $T_{0}-$ $\nabla _{x}T\cdot x_{0}.$ At
elevated currents, the Hall resistance $\rho _{xy}$ becomes dependent on
local temperature, hence, on sample width $d$ and current $I$. The thinner
is the sample and (or) higher the current, the higher is the local
temperature, thus, broader the interplateau transition. This mechanism
appeared to play some role in current-polarity dependent QHE breakdown.$^{%
\cite{Komiyama}}$

Let us estimate the external cooling of the contacts caused by 3D heat
leakage. For simplicity, we assume the high field corners(see Fig.(\ref
{Fig.1}), insert a) as hot(cold) points of radii $a$. These points are
placed into infinite 3D medium. For temperature difference $\Delta T$ the
total heat flux is of the order $\varkappa _{3D}a\Delta T$, where $\varkappa
_{3D}$ is the 3D thermal conductivity. This flux is less than the downstream
2DEG heat flux $\alpha TI$ when $\varkappa _{3D}al/(\sigma _{yx}LTd)=p<1$.
At $T=0.1$K the phonon related thermal conductivity for GaAs based structure
is $5\cdot 10^{-4}$W/(Km).$^{\cite{Ruf}}$ Assuming the point size being of
the order of 2DEG thickness $a\sim 30A$, then $l/d=3$, $\rho _{yx}\simeq
2/5h/e^{2}$ one obtains $p\simeq 2$. We stress, the above estimation gives
the upper limit for 3D phonon cooling. Actually, the cooling is much lower
because of finite size of sample.

We remind, up to now dc measurements case was discussed. However, our
approach could be applied as well for ac case . As was shown in Refs.\cite
{Kirby},\cite{Cheremisin}, at $B=0$ the Peltier effect related resistivity
vanishes above certain frequency dependent on inertial processes of thermal
diffusion. We recall, in quantizing magnetic field the microscopic current
and heat fluxes known to depend on the magnetism of the conducting
electrons. Actually, Eqs.(\ref{current+heat}) describe just average current
and heat densities of a bounded topology sample. We argue that diamagnetic
surface currents could define the dynamics of thermal processes in 2DEG.
Assuming $m=0.069m_{0}$ one could estimate the cyclotron frequency $\omega
_{c}=2.8\cdot 10^{13}$Hz and magnetic length $l_{H}=\left( \hbar /m\omega
_{c}\right) ^{1/2}=70$A at $B$=10T. For typical sample size $\sim 1$ mm the
transit time of an electron scattered at the sample boundaries is given $%
t\sim l/(\omega _{c}l_{H})=5\cdot 10^{-9}c$. Accordingly, one could estimate
the critical frequency $f_{c}\sim 1/t=0.2$GHz for thermal dynamics limit.

In conclusion, in strong quantum limit for dissipationless 2DEG(2DHG) under
current carrying conditions there exists effectively the longitudinal
resistivity caused by Peltier effect. The current results in
heating(cooling) at first(second) sample contact due to Peltier effect. At
small current the contacts temperatures are different, the temperature
gradient is linear on current. The voltage swing downstream the current is
proportional to Peltier effect induced thermopower. As a result, nonzero
longitudinal resistivity is measured in experiment. This value could be
erroneously identified as bulk resistivity of 2D electron gas. The above
effect could exists as well in 3D case.

\subsection{Appendix}

For 2DEG within zero approximation with respect to scattering the force, $%
{\bf F\perp B}$ results in macroscopic drift of electrons with the drift
velocity ${\bf v}_{dr}=-\frac{c}{eB^{2}}\left[ {\bf F}\times {\bf B}\right]
. $ The dissipasionless current and entropy densities are:$^{\cite{Zyryanov2}%
}$

\begin{eqnarray}
{\bf j} &=&-eN{\bf v}_{d}=\frac{c}{B^{2}}\left[ N{\bf F}\times {\bf B}%
\right] ,  \eqnum{A-1}  \label{App_I} \\
{\bf j}_{s} &=&sN{\bf v}_{d}=-\frac{c}{eB^{2}}\left[ S{\bf F}\times {\bf B}%
\right] ,  \nonumber
\end{eqnarray}
where $N=-\left( \frac{\partial \Omega }{\partial \mu }\right) _{T,B}$ , $%
S=-\left( \frac{\partial \Omega }{\partial T}\right) _{\mu ,B}$ are 2DEG
concentration and entropy densities respectively, $s=S/N$ is the entropy per
one electron. Then, ${\bf F=}-\nabla \zeta $ for nonuniform electrochemical
potential in x-y plane. The macroscopic current and heat flux density, ${\bf %
q}_{h}=T{\bf j}_{s}={\bf q}+{\bf j}\zeta /e$, are given 
\begin{eqnarray}
{\bf j} &=&-\frac{ceN}{B^{2}}\left[ {\bf E}\times {\bf B}\right] -\frac{cS}{%
B^{2}}\left[ {\bf \nabla T}\times {\bf B}\right] ,  \eqnum{A-2}
\label{App_II} \\
{\bf q}_{h} &=&\frac{cTS}{B^{2}}\left[ {\bf E}\times {\bf B}\right] +\frac{c%
}{eB^{2}}\frac{\partial }{\partial T}T^{2}\frac{\partial }{\partial T}\left[
\int\limits_{-\infty }^{\mu }\frac{\Omega }{T}d\mu ^{\prime }\right] \left[ 
{\bf \nabla T}\times {\bf B}\right] .  \nonumber
\end{eqnarray}
Eq.(\ref{App_II}) coincide with Eq.(\ref{current+heat}) Note, Eq.(\ref
{App_II}) could be generalized as$^{\cite{Zyryanov1}}$

\begin{eqnarray}
{\bf j} &=&\frac{c}{B^{2}}\text{rot}((Ne\varphi +\Omega ){\bf B)}, 
\eqnum{A-3}  \label{App_III} \\
{\bf q}_{h} &=&-\frac{c}{eB^{2}}T\left( \frac{\partial }{\partial T}\right)
_{\mu }\int\limits_{-\infty }^{\mu }\text{rot}((Ne\varphi +\Omega ){\bf B)}%
d\mu  \nonumber
\end{eqnarray}

\subsection{Appendix}

We now find out asymptotic equations for 2D thermopower and conductivity $%
\sigma _{yx}$ within low temperature and magnetic field limit $\nu ^{-1},\xi
<1$. Using well known Poisson formulae

\begin{equation}
\sum\limits_{n_{0}}^{\infty }\varphi (n)=\int\limits_{a}^{\infty }\varphi
(n)dn+2\text{Re}\sum\limits_{k=1}^{\infty }\int\limits_{a}^{\infty }\varphi
(n)e^{2\pi ikm}dm,  \eqnum{B-1}  \label{Poisson}
\end{equation}
where $m_{0}-1<a<m_{0}$, the 2DEG thermodynamic potential $\Omega (\mu ,\xi
,\nu )$ is given

\begin{equation}
\Omega =n_{0}\mu \xi ^{2}\left( -F_{1}(1/\xi )+2\pi
^{2}\sum\limits_{k=1}^{\infty }\frac{(-1)^{k}\cos (2\pi k\nu )}{r_{k}\sinh
(r_{k})}\right) .  \eqnum{B-2}  \label{Omega}
\end{equation}
Here, $n=\frac{m}{2\pi \hbar ^{2}}\mu $ is the 2DEG concentration at ${\bf B}%
=0$ (spin is neglected), $r_{k}=2\pi ^{2}\xi \nu k\sim kT/\hbar \omega _{c}$
is the dimensionless parameter. Therefore, $F_{n}(y)=\int\limits_{0}^{\infty
}x^{n}(1+\exp (x-y))^{-1}dx$ is the Fermi integral. Finally, 2DEG entropy $S$%
, concentration $N$ and are 
\begin{eqnarray}
N &=&n\xi \left( F_{0}(1/\xi )+2\pi \sum\limits_{k=1}^{\infty }\frac{%
(-1)^{k}\sin (2\pi k\nu )}{\sinh (r_{k})}\right) ,  \eqnum{B-3}  \label{N&S}
\\
S &=&kn\left( \frac{d}{d\xi }(F_{1}(1/\xi )\xi ^{2})-2\pi ^{2}\xi
\sum\limits_{k=1}^{\infty }(-1)^{k}\Phi (r_{k})\cos (2\pi k\nu )\right) , 
\nonumber
\end{eqnarray}
where $\Phi (z)=\frac{1-z\coth (z)}{z\cdot \sinh (z)}$. From Eq.(\ref{N&S}),
one could find out $\alpha =-S/eN$ and $\sigma _{yx}=Nec/B$.

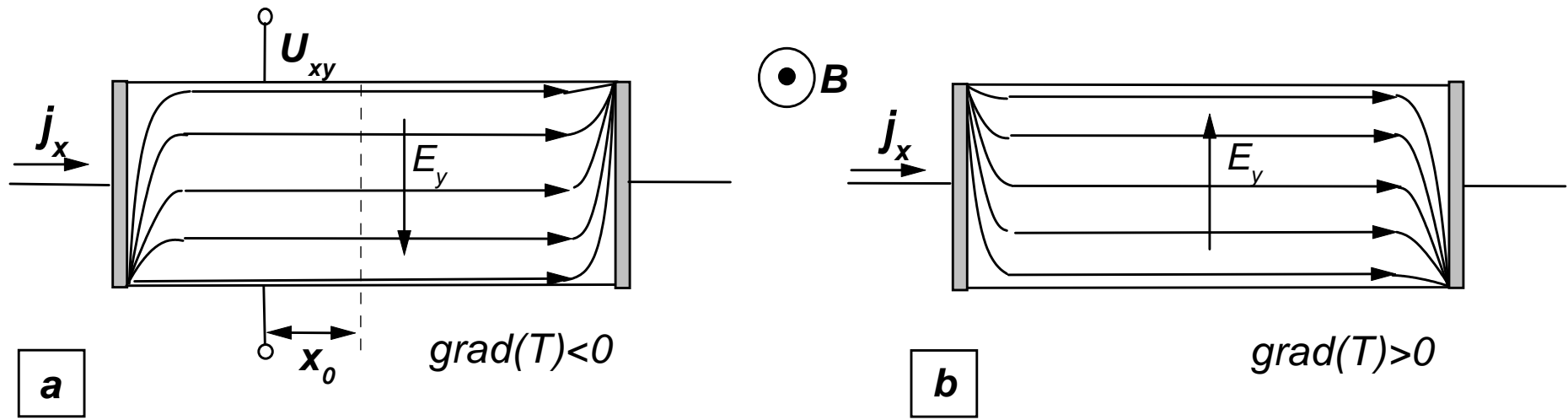
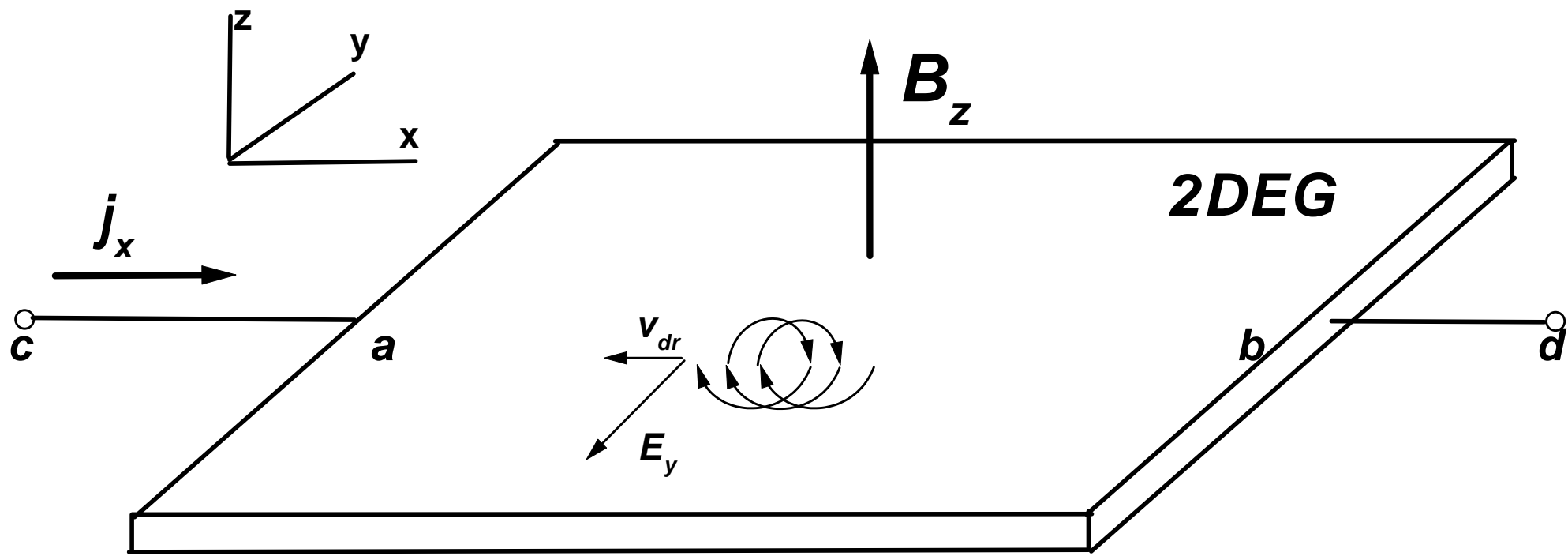
\begin{figure}[tbp]
\caption{Quantum Hall experimental setup. Inserts: current flow in a) 2DEG
b) 2DHG Hall bar samples.}
\label{Fig.1}
\end{figure}
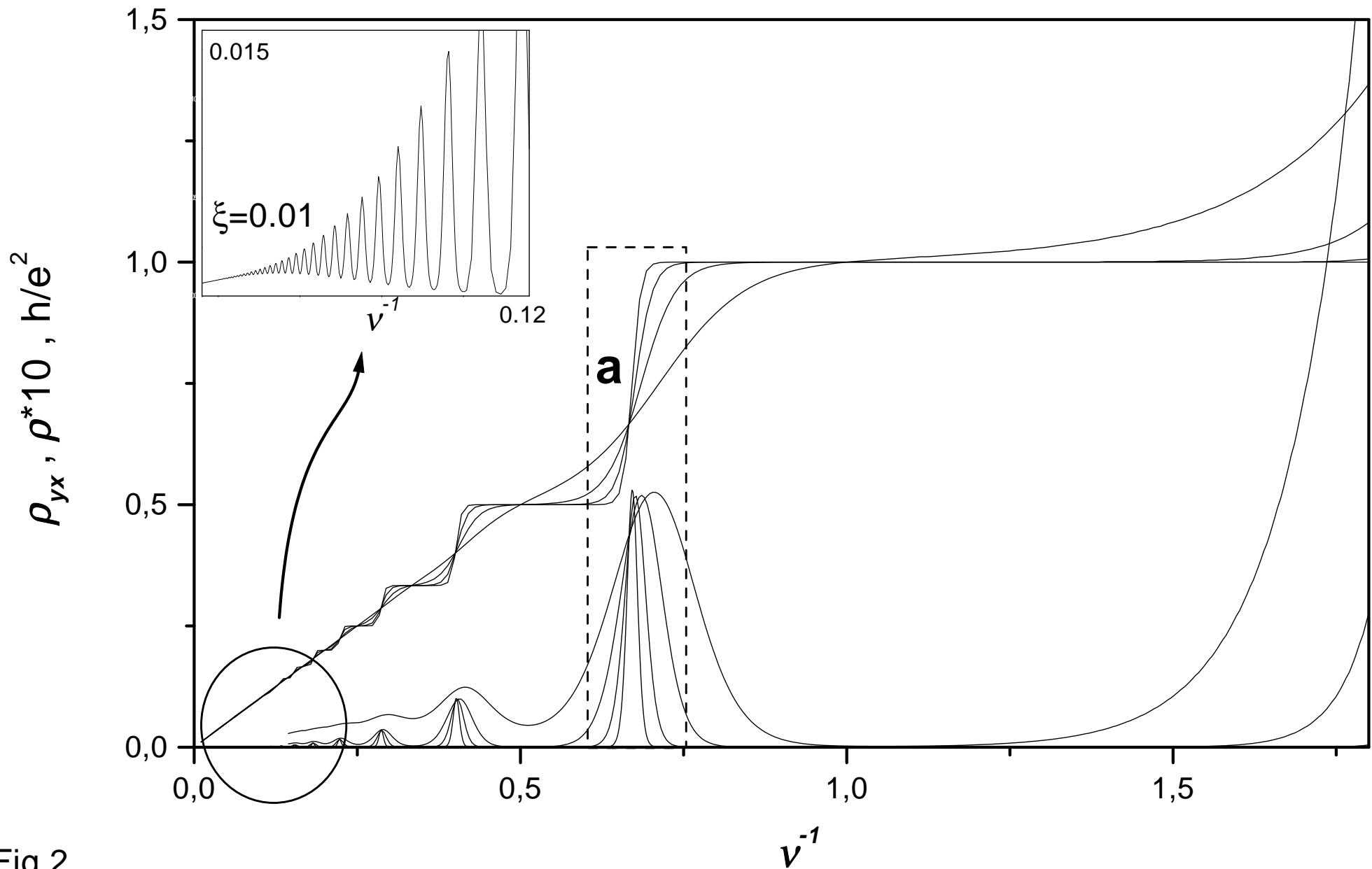
\begin{figure}[tbp]
\caption{Transverse magnetoresistance $\rho _{yx}$ and thermal correction $%
\rho $ (scaled by factor 10) given by Eq.(\ref{CorrectionB}) vs $\nu^{-1} $
for $\xi =0.01,0.02,0.04, 0.08$. Insert: Low field dependence $\rho
(\nu^{-1} ) $ for $\xi =0.01$ found out by means of Eqs.(\ref{N&S}) }
\label{Fig.2}
\end{figure}
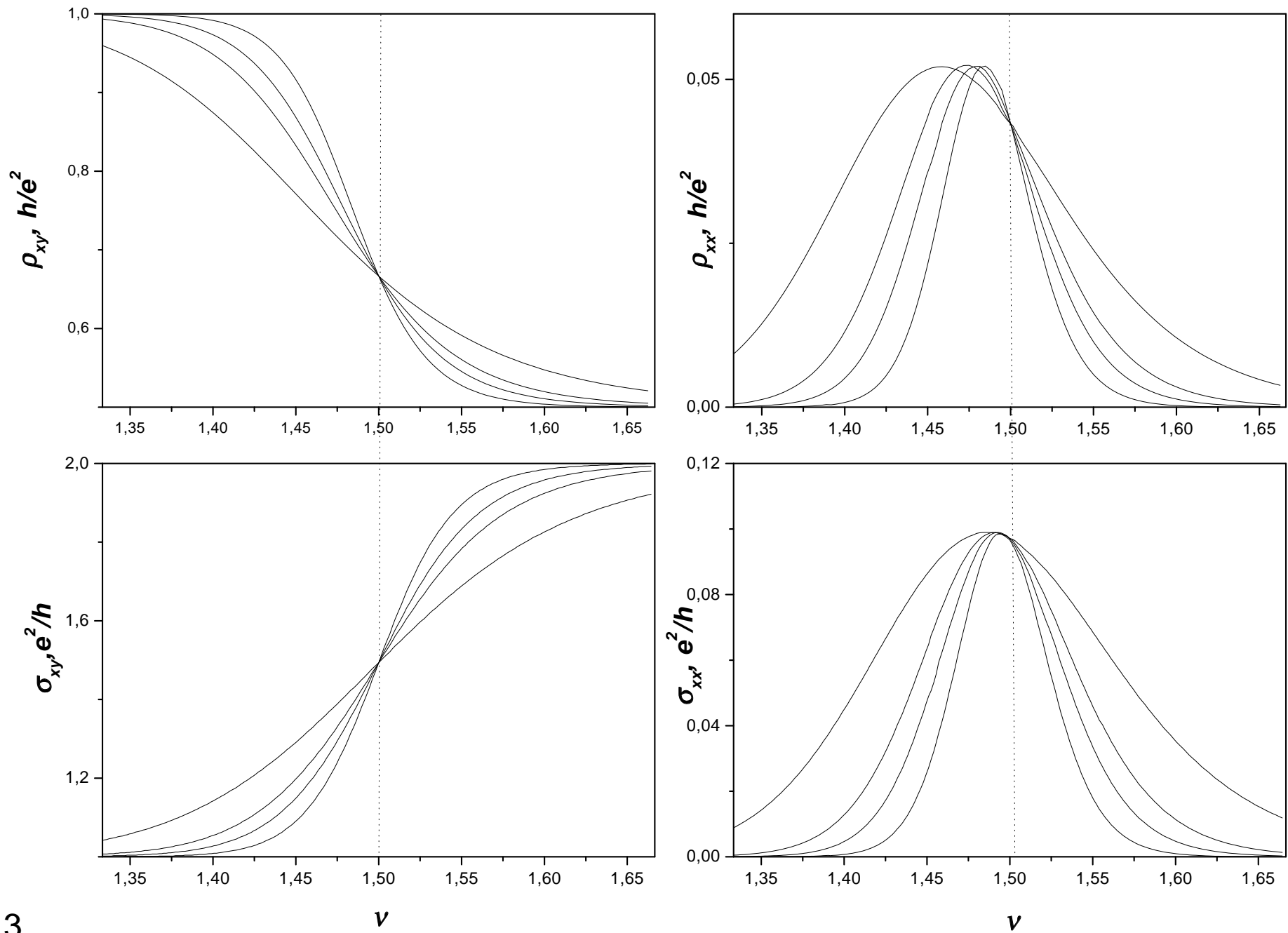
\begin{figure}[tbp]
\caption{Transverse magnetoresistance $\rho _{yx}$, thermal correction $\rho 
$ and conductancies $\sigma _{yx}^{*}$, $\sigma _{xx}^{*}$ nearby
half-filled LL $\nu_{c} =3/2$ (enlarged section (a) in Fig.(\ref{Fig.2}))
for $\xi =0.015,0.02,0.025,0.04$}
\label{Fig.3}
\end{figure}

\acknowledgments

The author is grateful to Prof. M.I.Dyakonov for criticism.

\end{document}